\documentclass{aastex631}
\usepackage{amsmath}
\usepackage{mathrsfs}
\received{2023 October 4}
\revised{2023 November 15}
\accepted{2023 November 16}
\published{2023 December 22}

\begin{document}

\title{Measuring Mass and Radius of the Maximum-mass Nonrotating Neutron Star}
\author[0000-0001-9120-7733]{Shao-Peng Tang}
\affiliation{Key Laboratory of Dark Matter and Space Astronomy, Purple Mountain Observatory, Chinese Academy of Sciences, Nanjing 210033, China}
\author{Bo Gao}
\affiliation{Key Laboratory of Dark Matter and Space Astronomy, Purple Mountain Observatory, Chinese Academy of Sciences, Nanjing 210033, China}
\affiliation{School of Astronomy and Space Science, University of Science and Technology of China, Hefei, Anhui 230026, China}
\author[0000-0001-5087-9613]{Yin-Jie Li}
\affiliation{Key Laboratory of Dark Matter and Space Astronomy, Purple Mountain Observatory, Chinese Academy of Sciences, Nanjing 210033, China}
\author[0000-0002-8966-6911]{Yi-Zhong Fan}
\correspondingauthor{Yi-Zhong Fan}
\email{yzfan@pmo.ac.cn}
\affiliation{Key Laboratory of Dark Matter and Space Astronomy, Purple Mountain Observatory, Chinese Academy of Sciences, Nanjing 210033, China}
\affiliation{School of Astronomy and Space Science, University of Science and Technology of China, Hefei, Anhui 230026, China}
\author[0000-0002-9758-5476]{Da-Ming Wei}
\affiliation{Key Laboratory of Dark Matter and Space Astronomy, Purple Mountain Observatory, Chinese Academy of Sciences, Nanjing 210033, China}
\affiliation{School of Astronomy and Space Science, University of Science and Technology of China, Hefei, Anhui 230026, China}

\begin{abstract}
The mass ($M_{\rm TOV}$) and radius ($R_{\rm TOV}$) of the maximum-mass nonrotating neutron star (NS) play a crucial role in constraining the elusive equation of state of cold dense matter and in predicting the fate of remnants from binary neutron star (BNS) mergers.
In this study, we introduce a novel method to deduce these parameters by examining the mergers of second-generation (2G) black holes (BHs) with NSs.
These 2G BHs are assumed to originate from supramassive neutron stars (SMNSs) formed in BNS mergers.
Since the properties of the remnant BHs arising from the collapse of SMNSs follow a universal relation governed by $M_{\rm TOV}$ and $R_{\rm TOV}$, we anticipate that by analyzing a series ($\sim 100$ detections) of mass and spin measurements of the 2G BHs using the third-generation ground-based gravitational-wave detectors, $M_{\rm TOV}$ and $R_{\rm TOV}$ can be determined with a precision of $\sim 0.01M_\odot$ and $\sim 0.6$ km, respectively.
\end{abstract}


\section{Introduction} \label{sec:intro}
The equation of state (EOS) of cold dense matter holds significant importance and is currently undetermined.
Observations of neutron stars (NSs), such as their masses and radii, can be utilized to constrain the unknown EOS.
The theoretical maximum mass ($M_{\rm TOV}$) that a nonrotating NS can sustain is a fundamental characteristic of an EOS.
The observed maximum mass of NSs through pulsar timing \citep[e.g., PSR J0740+6620,][]{2021ApJ...915L..12F} or the maximum cutoff mass inferred from NS population \citep{2018MNRAS.478.1377A,2020PhRvD.102f3006S,2023arXiv230912644F} serve as a stringent lower limit for $M_{\rm TOV}$, which can effectively eliminate various proposed EOS models.
While the (multimessenger) analyses of the NS merger-driven events, in particular GW170817/GRB 170817A/AT2017gfo, have yielded a rough estimate of $M_{\rm TOV}$ \citep{2013PhRvD..88f7304F, 2020PhRvD.101f3029S} or an upper limit \citep[e.g., ][]{2017ApJ...850L..19M, 2017PhRvD..96l3012S, 2018ApJ...852L..25R, 2018PhRvD..97b1501R} that rules out very stiff EOS.
If one has a series of mass-radius (or tidal deformability) measurements of NSs, the EOS can be determined by solving the so-called inverse stellar problem \citep{2014PhRvD..89f4003L}.\footnote{In other words, by varying the boundary conditions (like the central energy densities of NSs), a specific EOS will uniquely determine a corresponding mass-radius curve. These mass-radius curves, provided by different EOSs, can subsequently be employed to match the mass-radius observations, thereby imposing constraints on the EOS.}
For some NSs within low-mass X-ray binary (LMXB) systems, their mass and radius can be measured through spectroscopic observations of thermonuclear (Type-I) X-ray bursts exhibiting photospheric radius expansion, or by observing the angular size of the NSs during periods of quiescent states \cite[see, e.g., ][for a comprehensive review]{2016ARA&A..54..401O}.
Additionally, mass and radius information can also be derived from X-ray timing observations using a novel pulse profile modeling of the hot-spot emission from the surfaces of rotating NSs \citep{2019ApJ...887L..26B}, a method considered more reliable than the former approach \citep{2016EPJA...52...63M}.
The simultaneous measurements of mass and radius using this method have been achieved by the Neutron star Interior Composition Explorer (NICER) mission for two NSs: the nearby isolated NS PSR J0030+0451 \citep{2019ApJ...887L..21R, 2019ApJ...887L..24M} and the millisecond pulsar PSR J0740+6620 \citep{2021ApJ...918L..27R, 2021ApJ...918L..28M} with the highest mass known.
In compact binary coalescing systems involving at least one NS, the imprints left by tidal effects \citep{2008PhRvD..77b1502F, 2012PhRvD..85l3007D} in the emitted gravitational waves (GWs) provide a new avenue to investigate the EOS \citep{2018PhRvL.121p1101A,2018PhRvL.121i1102D}.
The first-ever measurements of mass and tidal deformability for a pair of coalescing NSs were obtained from the binary neutron star (BNS) merger event GW170817 \citep{2019PhRvX...9a1001A}, and subsequently other measurements were from GW190425 \citep{2020ApJ...892L...3A}.

The merger of a BNS system can lead to multiple outcomes, depending mainly on the total (gravitational) mass of the system ($M_{\rm tot}$) and the maximum mass of a nonrotating NS.
We briefly summarize the diverse outcomes as follows \cite[see, e.g., ][for a comprehensive review]{2017RPPh...80i6901B}:
(1) If the baryonic mass of the remnant ($M_{\rm b,rem}$) is less than that of the maximum-mass nonrotating NS ($M_{\rm b,TOV}$), it will initially form a rapidly rotating NS and eventually evolve into a stable, slowly rotating NS.
(2) When the gravitational mass of the remnant ($M_{\rm rem}$) is below the critical maximum mass ($M_{\rm crit}$) supported by uniform rotation, a supramassive neutron star (SMNS) will form. Subsequently, it will lose angular momentum \citep[or accrete matter from the envelope; ][]{2022ApJ...939...51M} and approach the stability line, leading to the formation of a rotating black hole (BH).
(3) If $M_{\rm rem}$ is slightly larger than $M_{\rm crit}$, a temporary existence of a hypermassive neutron star (HMNS) that can sustain more mass than uniform rotation becomes highly probable. After dissipative effects remove the differential rotation, it will become unstable and collapse, transforming either into an SMNS or directly into a rotating BH.
(4) In cases where $M_{\rm tot}$ exceeds a certain threshold for prompt collapse ($M_{\rm thres}$), a highly spinning BH (with a dimensionless angular momentum $\chi \simeq 0.7-0.8$, \citealt{2013PhRvD..88b1501K, 2014PhRvD..89j4021B}) can immediately form after the merger.
In the second scenario, it has been found that, at the time of collapse, the gravitational mass $M_{\rm rem}$ and the dimensionless angular momentum $\chi_{\rm rem}$ of the remnant satisfy a tight universal relation \citep{2020PhRvD.101f3029S}
\begin{equation}
\label{eq:univ-relation}
M_{\rm crit} = (1+c_2\mathscr{C}_{\rm TOV}^{-1}\chi_{\rm rem}^2+c_4\mathscr{C}_{\rm TOV}^{-2}\chi_{\rm rem}^4)M_{\rm TOV},
\end{equation}
where $M_{\rm crit}=M_{\rm rem}$, $c_2 = 0.0902(2)$, $c_4 = 0.0193(2)$, and $\mathscr{C}_{\rm TOV}=GM_{\rm TOV}/R_{\rm TOV}c^2$ ($G$ and $c$ are the gravitational constant and the speed of light, respectively) represents the compactness, which is associated with the mass and radius of a nonrotating NS at the maximum mass configuration.
This relation also holds for the mass ($M_{\rm BH}$) and spin ($\chi_{\rm BH}$) of the leftover BH, due to negligible GW radiation \citep{2023arXiv230606177D} and angular momentum conservation during the collapse.
Such a low-mass second-generation (2G) BH has the potential to encounter another NS, forming a binary system and possibly being detectable by GW detectors \citep{2020PhRvD.101j3036G,2021MNRAS.502.2049L,2023arXiv230709097G}.
If the mass and spin of the 2G BH can be precisely/accurately measured through the GW signal, a series of such measurements could lead to the determination of $M_{\rm TOV}$ and $R_{\rm TOV}$ by using the universal relation, which is the main motivation of this work.

\section{Methods}\label{sec:methods}
\begin{figure*}
    \centering
    \includegraphics[width=0.6\textwidth]{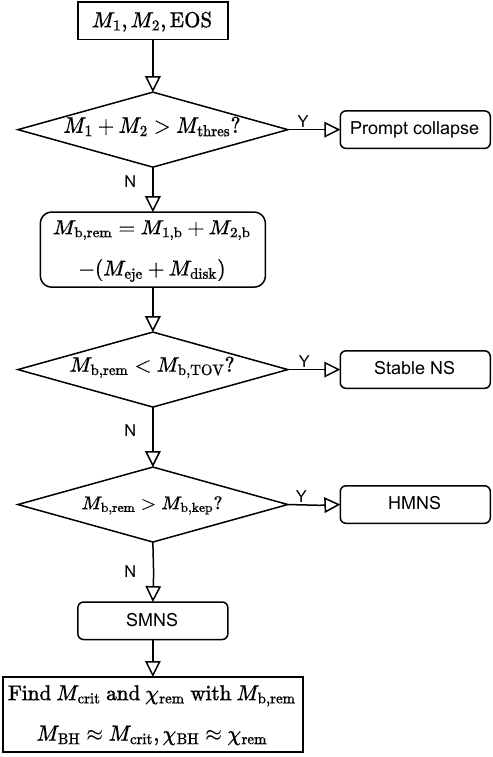}
    \caption{Flowchart for determining the properties of the final outcome resulting from the merger of a BNS system.}
    \label{fig:flowchart}
    \hfill
\end{figure*}
To illustrate our approach for constraining $M_{\rm TOV}$ and $R_{\rm TOV}$, we first select BSk21 \citep{2010PhRvC..82c5804G, 2013A&A...560A..48P} as the underlying EOS due to its good agreement with current observational constraints \citep[see, e.g.,][]{2021PhRvD.104f3032T}.
Then, we carry out the calculation of the stability line ($M_{\rm crit}-\chi$) for the given EOS using the Rapidly Rotating Neutron Star ({\sc RNS}) code \citep{1994ApJ...422..227C, 1995ApJ...444..306S, 1996PhDT.........4S}, which involves the following two steps:
(1) Compute the static and Keplerian sequences to determine their respective maximum (critical) masses (including, e.g., $M_{\rm TOV}$, $M_{\rm b,TOV}$, and $M_{\rm b,kep}$), along with other associated quantities, such as the central energy densities $\epsilon_{\rm c,TOV}$ and $\epsilon_{\rm c,kep}$, as well as the angular momentum $J_{\rm kep}$.
(2) Slice the identified $J_{\rm kep}$ and apply the `fminbound' algorithm from `scipy.optimize' with a boundary of ($\epsilon_{\rm c,kep}$, $\epsilon_{\rm c,TOV}$) to precisely locate the turning point within the J-constant model.
Through analyses of NS mass distribution and EOS-insensitive relations, \citet{2020PhRvD.102f3006S} found that the prospect of forming SMNS during BNS mergers is quite promising \citep[see also][for the latest validation]{2023arXiv230912644F}.
\begin{table*}
\centering
\caption{Mass Measurements of BNS Systems and Predicted Remnant's Properties (based on EOS BSk21)}
\label{tb:metadata}
\begin{tabular}{lccccccccc}
\hline
Source    & $M_1$    & $M_2$    & Reference    & $M_{\rm thres}$    & $M_{\rm 1,b}$    & $M_{\rm 2,b}$    & $M_{\rm eje}+M_{\rm disk}$    & $M_{\rm rem}$    & $\chi_{\rm rem}$ \\ \hline
B1534+12    & $1.3455$    & $1.333$    & \citet{2014ApJ...787...82F}    & $3.09$    & $1.47$    & $1.46$    & $0.07$    & $2.41$    & $0.42$ \\
B1913+16    & $1.4398$    & $1.3886$    & \citet{2010ApJ...722.1030W}    & $3.08$    & $1.59$    & $1.52$    & $0.05$    & $2.59$    & $0.61$ \\
B2127+11C    & $1.358$    & $1.354$    & \citet{2006ApJ...644L.113J}    & $3.09$    & $1.49$    & $1.48$    & $0.05$    & $2.46$    & $0.48$ \\
J0453+1559    & $1.559$    & $1.174$    & \citet{2015ApJ...812..143M}    & $2.88$    & $1.74$    & $1.27$    & $0.18$    & $2.37$    & $0.36$ \\
J0509+3801    & $1.46$    & $1.34$    & \citet{2018ApJ...859...93L}    & $3.05$    & $1.61$    & $1.47$    & $0.09$    & $2.52$    & $0.55$ \\
J0514-4002A    & $1.25$    & $1.22$    & \citet{2019MNRAS.490.3860R}    & $3.08$    & $1.36$    & $1.32$    & $0.22$    & $\approx 2.11$    & $<0.05$ \\
J0737-3039    & $1.338185$    & $1.248868$    & \citet{2021PhRvX..11d1050K}    & $3.06$    & $1.46$    & $1.36$    & $0.14$    & $\approx 2.25$    & $<0.05$ \\
J1756-2251    & $1.341$    & $1.23$    & \citet{2014MNRAS.443.2183F}    & $3.05$    & $1.47$    & $1.33$    & $0.16$    & $\approx 2.23$    & $<0.05$ \\
J1757-1854    & $1.3922$    & $1.3406$    & \citet{2023mgm..conf.3774C}    & $3.08$    & $1.53$    & $1.47$    & $0.07$    & $2.47$    & $0.49$ \\
J1807-2500B    & $1.3655$    & $1.2064$    & \citet{2012ApJ...745..109L}    & $3.03$    & $1.50$    & $1.31$    & $0.17$    & $\approx 2.22$    & $<0.05$ \\
J1906+0746    & $1.322$    & $1.291$    & \citet{2015ApJ...798..118V}    & $3.08$    & $1.44$    & $1.41$    & $0.11$    & $2.30$    & $0.16$ \\
J1829+2456    & $1.306$    & $1.299$    & \citet{2021MNRAS.500.4620H}    & $3.09$    & $1.42$    & $1.42$    & $0.10$    & $2.29$    & $0.14$ \\
J1913+1102    & $1.62$    & $1.27$    & \citet{2020Natur.583..211F}    & $2.93$    & $1.81$    & $1.38$    & $0.18$    & $2.55$    & $0.58$ \\
GW170817    & $1.46$    & $1.27$    & \citet{2019PhRvX...9c1040A}    & $3.01$    & $1.62$    & $1.38$    & $0.13$    & $2.42$    & $0.44$ \\
GW190425    & $1.4$    & $1.95$    & \citet{2021PhRvX..11b1053A}    & $3.11$    & $1.54$    & $2.25$    & $\approx 0$    & $\approx 3.35$    & $\gtrsim 0.7$ \\ \hline
\multicolumn{10}{l}{Notes. All the masses are measured in units of solar mass.} \\
\multicolumn{10}{l}{The $M_{\rm b,TOV}$, $M_{\rm b,kep}$, and $\chi_{\rm kep}$ for EOS BSk21 are $2.72M_\odot$, $3.22M_\odot$, and $0.7$, respectively.}
\end{tabular}
\end{table*}
Here we perform calculations to determine the properties of the final outcome resulting from the merger within the presently identified BNS population (see Table~\ref{tb:metadata} for details), while considering the specified EOS.
The brief steps are summarized in the flowchart depicted in Fig.~\ref{fig:flowchart}.
First, we calculate the threshold mass for prompt collapse ($M_{\rm thres}$) using the fitting formula provided in \citet{2021ApJ...922L..19T} (see also \citealt{2019ApJ...872L..16K} and \citealt{2021PhRvD.103l3004B}).
If the total gravitational mass of the binary is below $M_{\rm thres}$, we then employ the empirical fitting formula $P_2^2(q,\tilde{\Lambda})$ (with recommended coefficients) from \citet{2022CQGra..39a5008N} to estimate the masses of both the dynamical ejecta ($M_{\rm eje}$) and the remnant disk ($M_{\rm disk}$).
Utilizing the conservation of baryonic mass within the system, we subsequently estimate the baryonic mass of the remnant $M_{\rm b,rem}=M_{\rm 1,b}+M_{\rm 2,b}-(M_{\rm eje}+M_{\rm disk})$, where $M_{\rm 1,b}$ and $M_{\rm 2,b}$ are component baryonic masses calculated with RNS.
If $M_{\rm b,rem} \in [M_{\rm b,TOV}, M_{\rm b,kep}]$, we proceed to determine the corresponding values of $M_{\rm crit}$ and $\chi$ for the given $M_{\rm b,rem}$ using the information obtained from the stability line established earlier.
If there are no measurement errors for $M_{\rm crit}$ and $\chi$, we can finally solve for the unknown values of $M_{\rm TOV}$ and $R_{\rm TOV}$ by fitting the data using the universal relation, i.e., Eq.~(\ref{eq:univ-relation}).
\begin{figure*}
    \centering
    \includegraphics[width=0.98\textwidth]{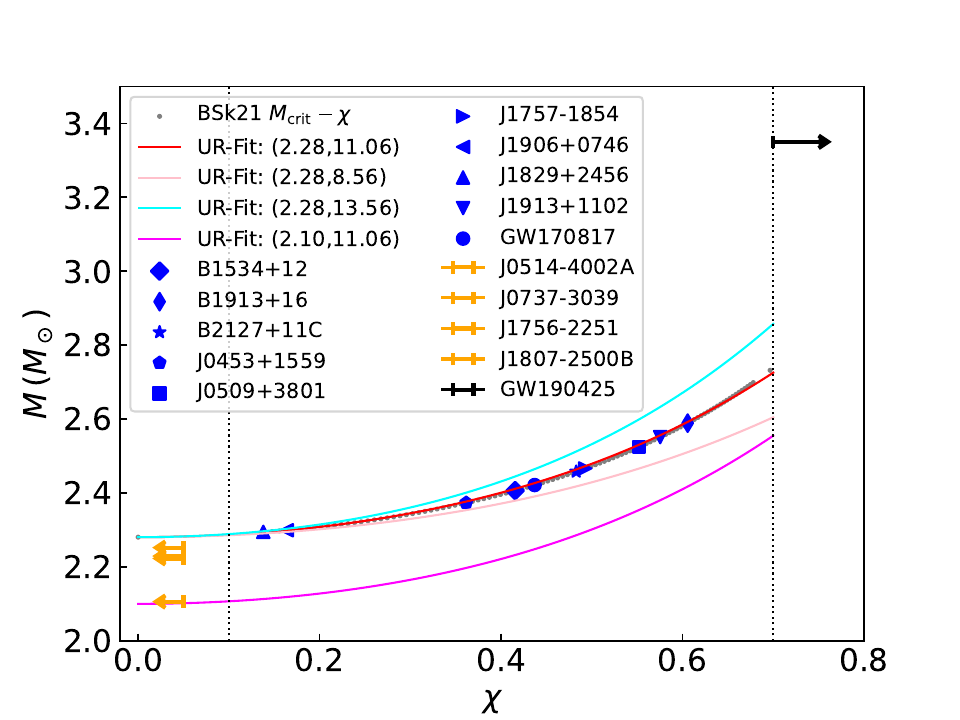}
    \caption{Stability line ($M_{\rm crit}-\chi$, depicted as a gray dotted line) of SMNS for the BSk21 EOS, along with the anticipated remnant masses and spins for BNS systems.
    These remnant characteristics can be divided into three categories, symbolized by blue, orange, and black markers.
    These markers correspond to SMNS remnants, stable NSs, and promptly formed BHs, respectively.
    The cyan, red, pink, and magenta lines illustrate curves based on universal relation, each with varying $M_{\rm TOV}$ and $R_{\rm TOV}$ (corresponding to values shown in the parentheses of the legend).
    The vertical black dotted lines mark the $\chi=0.1$ and $\chi=0.7$ boundaries, which roughly separate the three categories.}
    \label{fig:bns-rem}
    \hfill
\end{figure*}

However, in reality, the measured values of $M_{\rm BH}$ (associated with $M_{\rm crit}$) and $\chi_{\rm BH}$ in the merger of a 2G BH with an NS are subject to uncertainties.
To investigate how well the mass and spin parameters can be measured for such BH-NS systems by future GW detectors, e.g., the Einstein Telescope \citep[ET;][]{2010CQGra..27s4002P} and the Cosmic Explorer \citep[CE;][]{2019BAAS...51g..35R}, we estimate the measurement uncertainties for a simulated population of these systems using a novel Fisher-matrix code {\sc GWFAST} \citep{2022ApJ...941..208I, 2022ApJS..263....2I} with  {\sc IMRPhenomXP\_NRTidalv2} waveform template \citep{2019PhRvD.100d4003D, 2021PhRvD.103j4056P}.
We assume that $\chi_{\rm BH}$ is uniformly distributed in the range of (0.1, 0.7), and $M_{\rm BH}$ is determined using Eq.~(\ref{eq:univ-relation}) with the true values of $M_{\rm TOV}$ and $R_{\rm TOV}$.
The spin of the NS is assumed to have a low magnitude, distributed uniformly between 0 and 0.05.
The NS mass follows a bimodal Gaussian distribution as defined in Eq.~(1) of \citet{2020PhRvD.102f3006S}, and is characterized by $M_{\rm min} = 0.9M_\odot$, $\mu_1 = 1.33M_\odot$, $\sigma_1 = 0.08M_\odot$, $\mu_2 = 1.79M_\odot$, $\sigma_2 = 0.31M_\odot$, $r = 0.57$, and a maximum-mass cutoff at $M_{\rm TOV} = 2.28M_\odot$.
Besides, it is assumed that both the spin orientations of the BH and the NS are isotropically distributed, and the azimuthal position of orbital angular momentum on its cone of precession about the total angular momentum $\phi_{\rm JL}$ and the azimuthal angle separating the spin vectors $\phi_{12}$ are uniform in (0,2$\pi$).
We employ the redshift distribution similar to that of \citet{2022ApJ...941..208I}, and with a longer minimum time delay $t_{\rm d,min} = 100$ Myr.
Additionally, we maintain the assumption of no redshift evolution in the mass distribution.
The luminosity distance is computed from the redshift assuming {\sc Planck}18 flat $\Lambda$CDM \citep{2020A&A...641A...6P}.
We then use the complete GW parameters (for a list and description of these parameters, one can refer, for example, to Table~1 of \citet{2020ApJ...892...56T}) to calculate the Fisher Information Matrix (FIM), except for fixing the tidal deformability of the BH to $0$.
We discard all events where the inversion error of the FIM exceeds $0.05$ and where the optimal signal-to-noise ratio (S/N) is below $12$.  
After obtaining the distributions of $\Delta M_{\rm BH}$ (source frame mass) and $\Delta \chi_{\rm BH}$, we then randomly assign measurement errors to a set of mock events.
For each event, we introduce random shifts drawn from normal distributions with standard deviations corresponding to the measurement errors.
These shifts are added to the actual values of $M_{\rm BH}$ and $\chi_{\rm BH}$, yielding the corresponding `measured' mean values.
This process is intended to simulate the potential effects encountered in real data analysis, where slight biases in parameter inference may arise.
Finally, we utilize the orthogonal distance regression (ODR) method implemented in the `scipy' library to perform data fitting.
This is specifically applied to the mock data $M_{\rm BH}$ and $\chi_{\rm BH}$, taking into consideration their respective measurement errors.
This approach allows us to determine the best-fit parameters for $M_{\rm TOV}$ and $R_{\rm TOV}$, along with their associated uncertainties.

\section{Results} \label{sec:results}
As illustrated in Fig.~\ref{fig:bns-rem}, based on the BSk21 EOS, the majority of Galactic double NS systems are expected to form SMNS if they were to merge.
The heaviest BNS system currently observed, GW190425, is associated with a high-spin value of the remnant, as indicated by numerical-relativity simulations \citep{2022PhRvD.106h4039D}.
In contrast, some lighter double NS systems are presumed to produce low-spin stable NSs.
The universal relation, characterized by the true $M_{\rm TOV}$ and $R_{\rm TOV}$ parameters of BSk21 (represented by the red curve), fits the `observed' remnant masses and spins well.
By varying these parameters, we notice that $M_{\rm TOV}$ has a strong influence on the overall data fit, while $R_{\rm TOV}$ predominantly impacts the fitting of high-spin data.

\begin{figure*}
    \centering
    \includegraphics[width=0.98\textwidth]{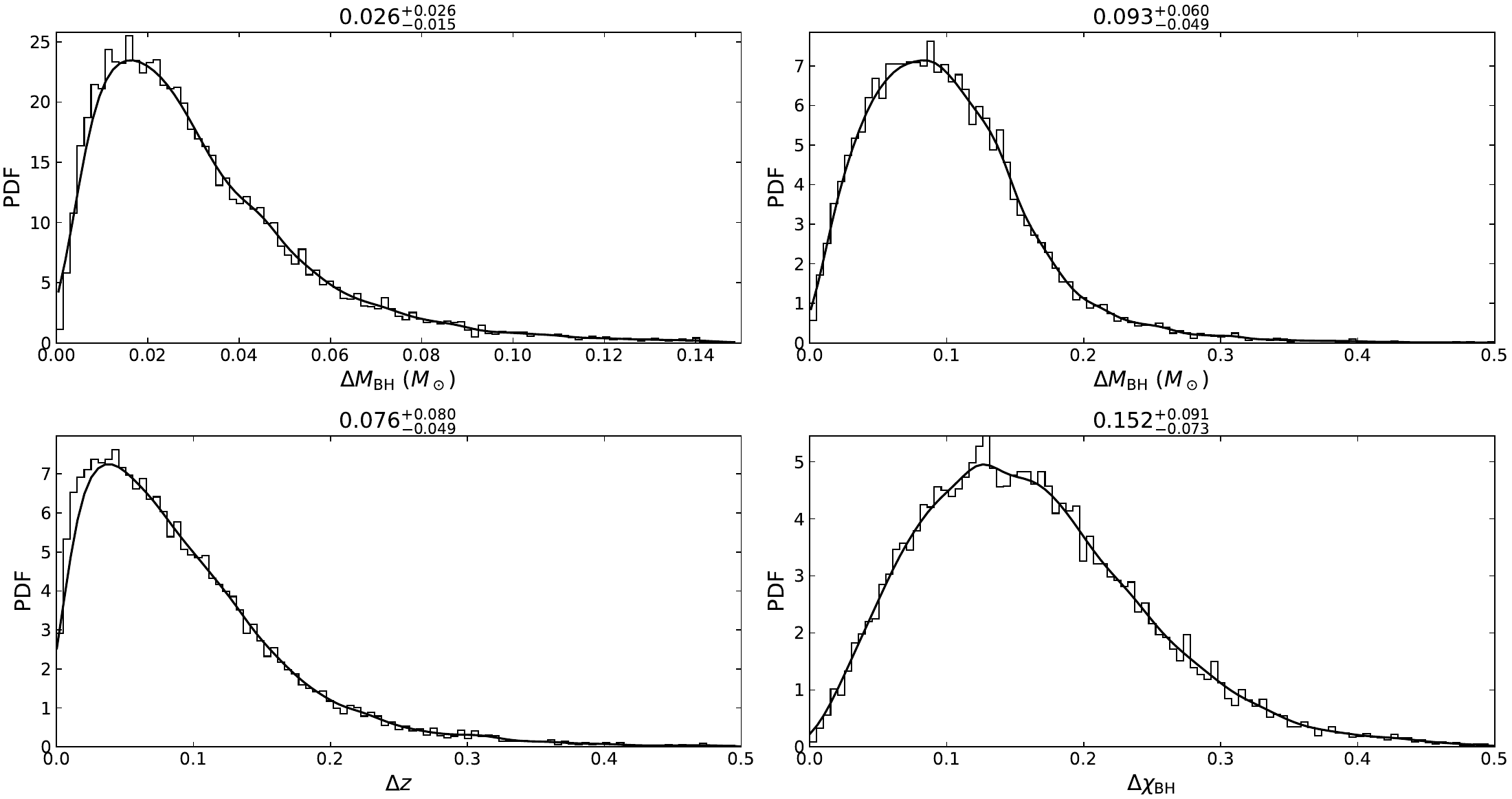}
    \caption{Distributions of measurement uncertainties for the redshift $z$ and the source frame mass $M_{\rm BH}$ and spin magnitude $\chi_{\rm BH}$ of a BH under the `ET+2CE' detector configuration.
    Top panels present the $\Delta M_{\rm BH}$ results with (left) and without (right) electromagnetic counterparts information.
    Bottom panels show the results of $\Delta z$ (left) derived from the uncertainty of luminosity distance and  the results of $\Delta \chi_{\rm BH}$ (right).
    The smooth, black solid lines represent the kernel density estimates (KDEs) of the distributions.
    The numbers reported above each panel are the median and $68.3\%$ symmetric credible interval.}
    \label{fig:fisher-res}
    \hfill
\end{figure*}
For (2G)BH-NS systems that are detectable (with S/N$\geq 12$) by future GW detectors, we showcase the distributions of measurement uncertainties for the source frame mass, $M_{\rm BH}$, and spin magnitude, $\chi_{\rm BH}$, of the BH in Fig.~\ref{fig:fisher-res}.
The detector configuration is consistent with the `ET+2CE' settings as used in \citet{2022ApJ...941..208I}.
We determine the measurement uncertainty of the source frame mass of BHs by
\begin{equation}
\Delta M_{\rm BH}=\sqrt{\left(\frac{\Delta M_{\rm BH}^{z}}{1+z}\right)^2 + \left(\frac{M_{\rm BH}^{z} \Delta z}{(1+z)^2}\right)^2},
\end{equation}
where $\Delta M_{\rm BH}^{z}$ is the measurement uncertainty of the detector frame mass of BHs.
The redshift uncertainty $\Delta z$ is calculated through
\begin{equation}
\Delta z=\Delta d_L/\left(\frac{d_L}{1+z}+\frac{c(1+z)}{H(z)}\right),
\end{equation}
where $H(z)$ is the Hubble parameter as a function of redshift.
As illustrated in the top panels of Fig.~\ref{fig:fisher-res}, it is evident that the proposed third-generation GW detectors exhibit significant capability in constraining the component masses.
For BH-NS mergers, there is the possibility of producing electromagnetic (EM) radiations, such as gamma-ray bursts (GRBs) and kilonovae.
This can help determine the redshift (we assume $\Delta z \sim 0$ in this case), consequently reducing the measurement uncertainty of $M_{\rm BH}$.
The $\Delta M_{\rm BH}$ derived solely from GW observations (hereafter referred to as `GW only') is approximately ${0.093}_{-0.049}^{+0.060}\,M_\odot$.
This capability is further enhanced with the assistance of EM counterparts (hereafter referred to as `GW+EM'), with $\Delta M_{\rm BH}={0.026}_{-0.015}^{+0.026}\,M_\odot$.
While the ability to constrain the spin magnitude is not as pronounced as for the mass, as shown in the bottom right panel of Fig.~\ref{fig:fisher-res}.

\begin{figure*}
    \centering
    \includegraphics[width=0.49\textwidth]{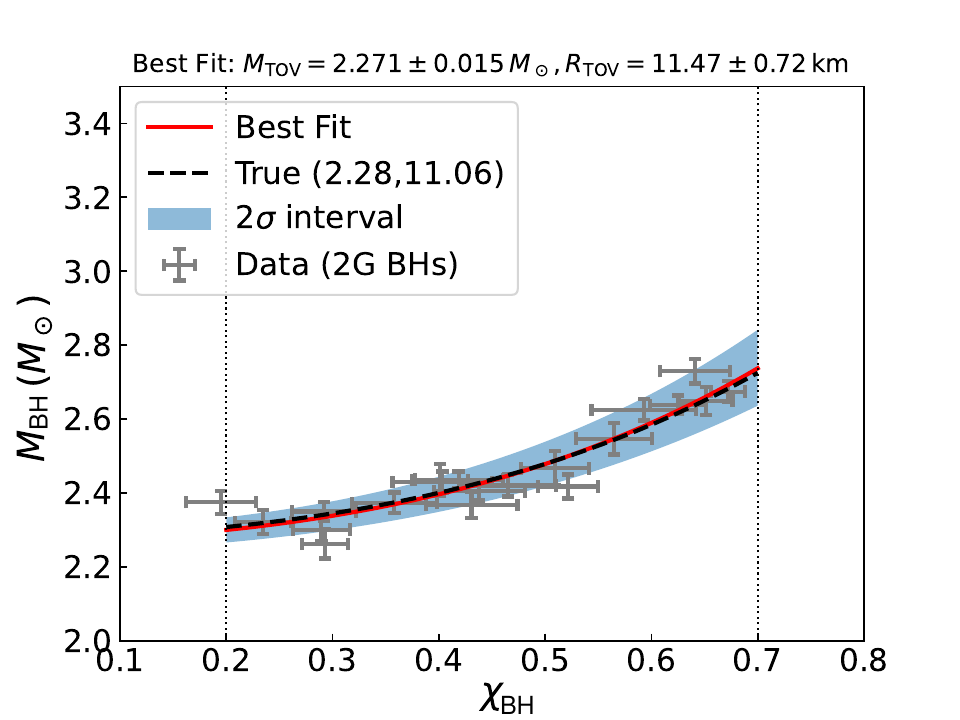}
    \includegraphics[width=0.49\textwidth]{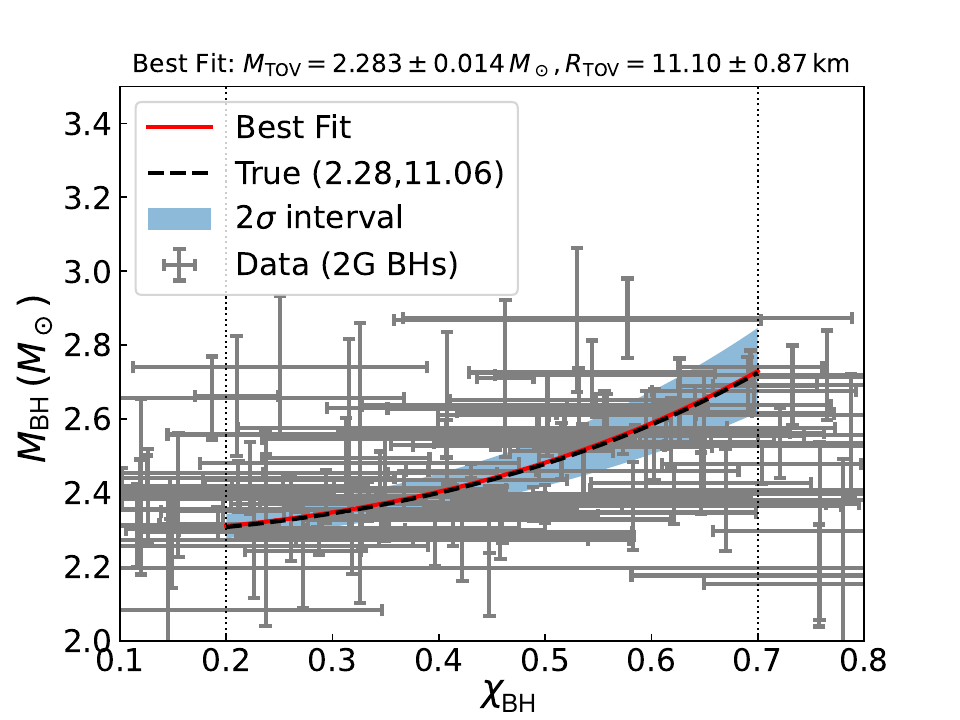}
    \includegraphics[width=0.49\textwidth]{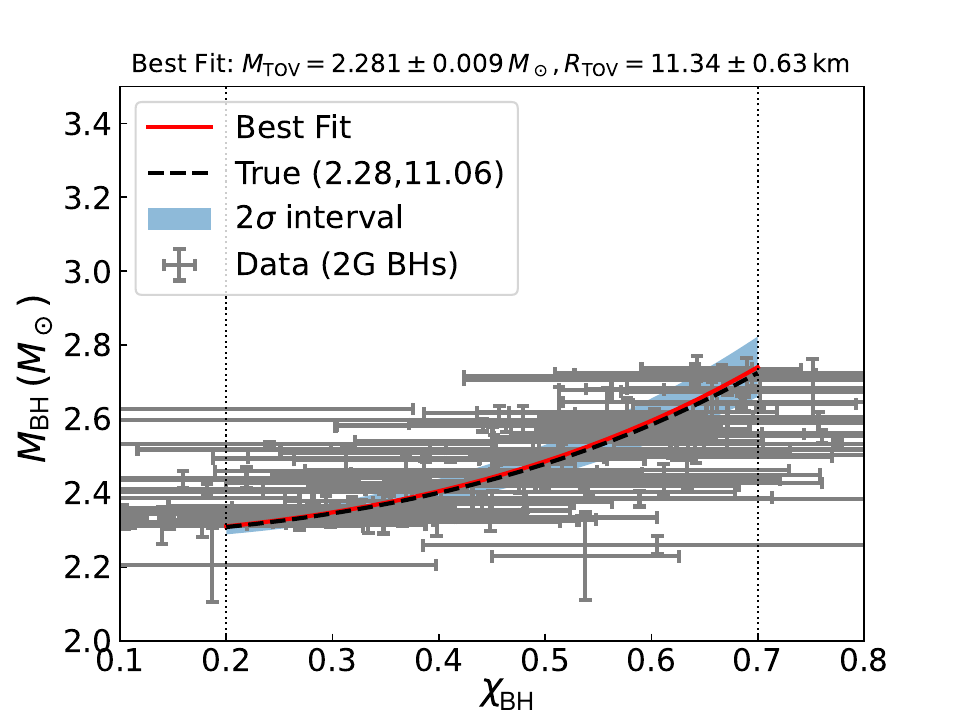}
    \includegraphics[width=0.49\textwidth]{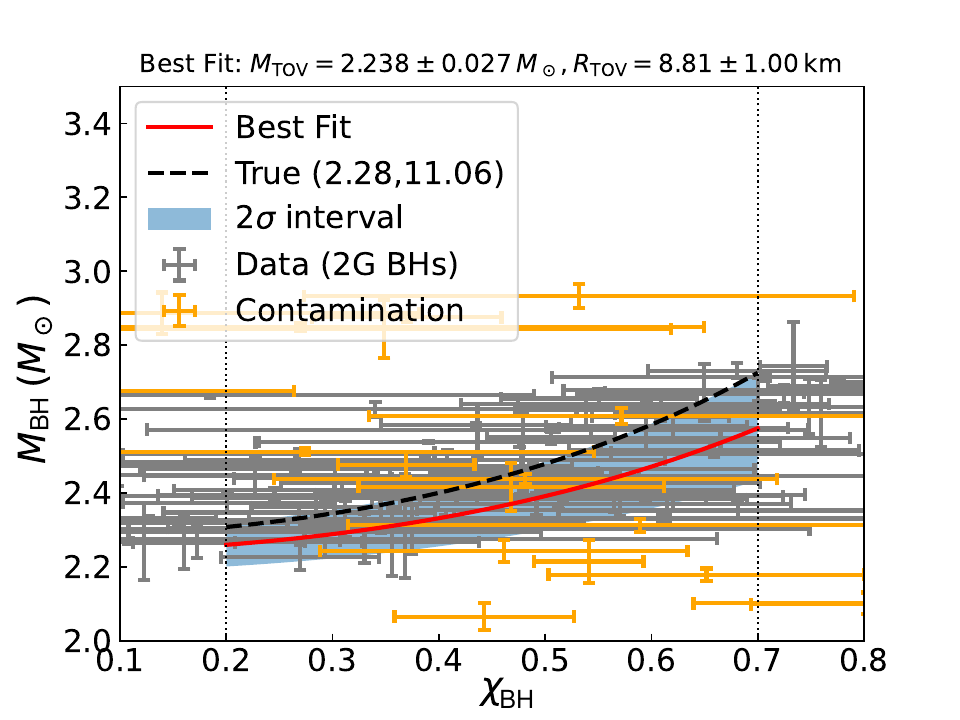}
    \caption{Results of ODR fitting on randomly generated measurement samples using the universal relation described by Eq.~(\ref{eq:univ-relation}).
    The top left panel serves as an illustration of the method, showcasing the fitting results with 20 golden events.
    The top right and bottom left panels exhibit the fitting results for 100 samples randomly generated from the distribution of $\Delta M_{\rm BH}$ for the `GW only' and `GW+EM' cases, respectively.
    The bottom right panel evaluates the effect of outlier contamination on the fitting procedure.
    In each panel, the red solid curve represents the universal relation with the best-fit parameter values, and the black dashed line represents that with the true values.
    The light blue band indicates the $2\sigma$ interval, and the gray error bars denote the mock measurements.
    The numbers shown above each panel represent the best-fit values and their corresponding $1\sigma$ uncertainties.}
    \label{fig:univ-fit}
    \hfill
\end{figure*}
\begin{figure*}
    \centering
    \includegraphics[width=0.8\textwidth]{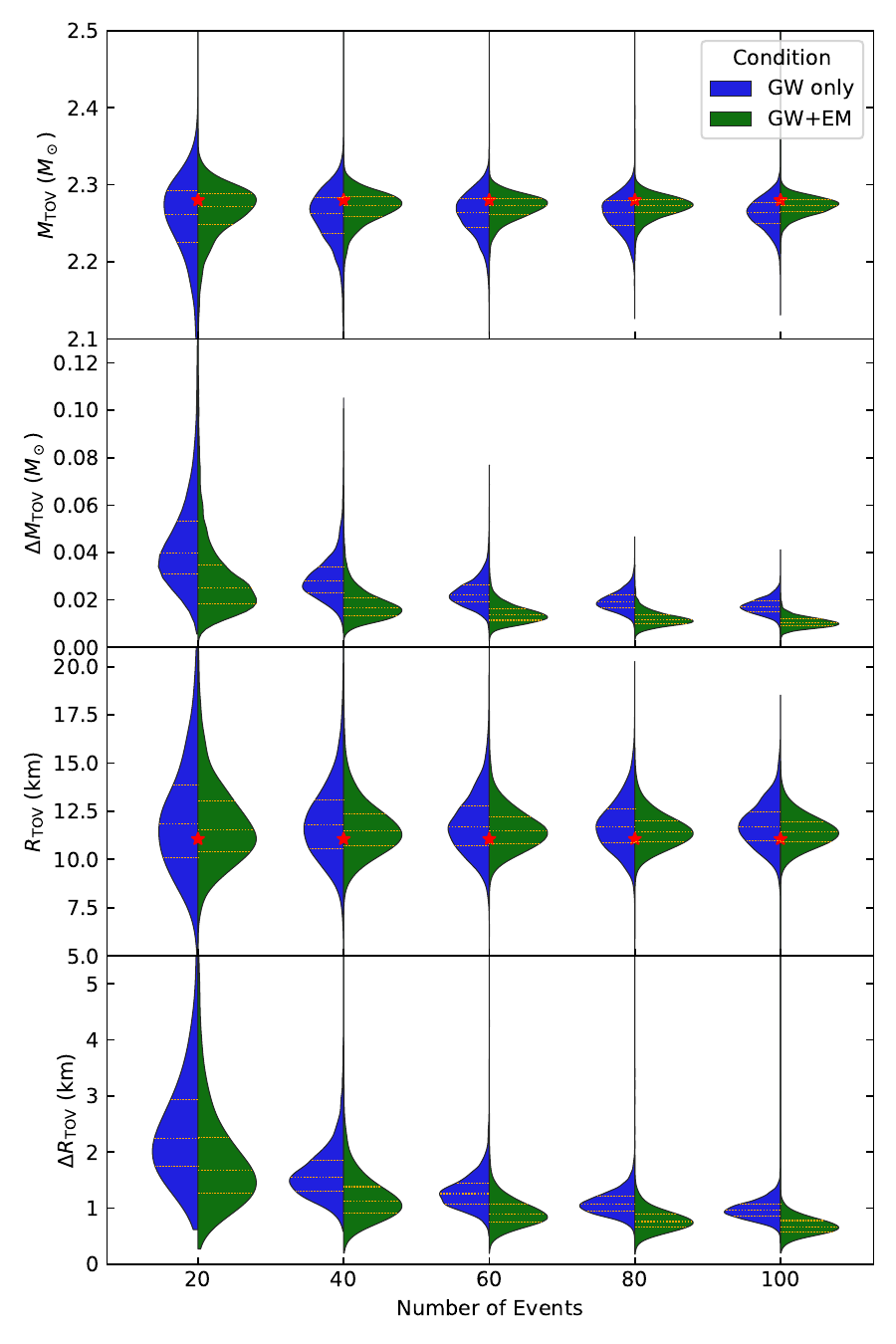}
    \caption{Distributions of the best-fit values and corresponding $1\sigma$ uncertainties of $M_{\rm TOV}$ and $R_{\rm TOV}$.
    The blue and green violin plots represent the `GW only' and `GW+EM' results, respectively.
    The orange dashed/dotted lines denote the median and the $68.3\%$ interval.
    The red star markers represent the true values.}
    \label{fig:ave-res}
    \hfill
\end{figure*}

With the distribution of measurement uncertainties in hand, we can proceed to generate mock measurements and employ ODR fitting to determine the values of $M_{\rm TOV}$ and $R_{\rm TOV}$.
To illustrate this method, we initially utilize a set of 20 `golden' events, each with precise measurements of $M_{\rm BH}$ and $\chi_{\rm BH}$ (specifically, both have measurement error values $\lesssim0.05$).
As shown in the top left panel of Fig.~\ref{fig:univ-fit}, it is feasible to determine the unknown parameters (i.e., $M_{\rm TOV}$ and $R_{\rm TOV}$), and their true values can be accurately recovered.
In principle, with precise measurements of just two samples of $M_{\rm BH}$ and $\chi_{\rm BH}$, we can solve for these two parameters.
Increasing the number of data points will effectively reduce the uncertainties of $M_{\rm TOV}$ and $R_{\rm TOV}$.
To simulate a realistic scenario, we generate samples based on the capabilities of third-generation GW detectors for constraining mass and spin parameters.
As shown in the top right panel of Fig.~\ref{fig:univ-fit}, we notice that 100 `GW only' detections can produce a precision of $0.014~M_\odot$ for $M_{\rm TOV}$ and $0.9$ km for $R_{\rm TOV}$.
Furthermore, when we include additional EM counterpart information (i.e., the `GW+EM' case), the uncertainties further decrease to $0.01~M_\odot$ for $M_{\rm TOV}$ and $0.6$ km for $R_{\rm TOV}$.
However, if there are potential contaminants from additional BH populations (e.g., low-mass 1G BHs with high spins produced by accretion), $M_{\rm TOV}$ and $R_{\rm TOV}$ may be biased as shown in the bottom right panel of Fig.~\ref{fig:univ-fit}.
Considering that there are fluctuations in the random process used to generate the mock events, we then simulate 5000 instances of event samples.
The comprehensive results are illustrated in Fig.~\ref{fig:ave-res}, showcasing the distributions of recovered best-fit values along with their corresponding uncertainties from the ODR fitting.
These results demonstrate that as the number of detected events increases, both $M_{\rm TOV}$ and $R_{\rm TOV}$ can be measured with good accuracy and precision.
In the optimistic scenario, the precision of determining $M_{\rm TOV}$ and $R_{\rm TOV}$ can reach approximately $0.01\,M_\odot$ and $0.6$ km, respectively.

\section{Summary and Discussion} \label{sec:discussion}
In this work, we present a method for determining the mass and radius of a maximum-mass nonrotating neutron star, which exploits the tight universal relation between the critical mass of a uniformly rotating SMNS and its dimensionless angular momentum at the point of collapse.
Such a universal relation is associated with the mass ($M_{\rm TOV}$) and radius ($R_{\rm TOV}$) of the maximum-mass nonrotating neutron star.
The characteristics of the BHs formed after the collapse of SMNSs also largely conform to this universal relation.
Assuming that the remnant BH has the opportunity to merge with another NS, and that this event is detectable by GW detectors, it is theoretically possible to determine the parameters within the universal relation by utilizing measurements of the mass and spin of the remnant BH.
To evaluate the feasibility of this approach, we initially assess the capability of the proposed third-generation GW detectors to measure the mass and spin parameters.
This is done using the Fisher-matrix approach, as implemented in the GWFAST code.
Our findings suggest that it is feasible to accurately measure the source frame BH mass, while the measurement of the spin parameter is satisfactory, though not excellent.
Following these findings, we use ODR fitting to analyze mock measurement samples, considering different numbers of detected events and the inclusion or exclusion of EM counterpart information.
We find that it is highly probable to accurately and precisely measure both the $M_{\rm TOV}$ and $R_{\rm TOV}$.
In scenarios with 100 detections that include both GW and EM counterpart information, the $M_{\rm TOV}$ and $R_{\rm TOV}$ can be measured with a precision of approximately $0.01\,M_\odot$ and $0.6$ km, respectively.
This precise determination of the $M_{\rm TOV}$ and $R_{\rm TOV}$ holds significant potential for constraining the unknown EOS of dense matter and for enhancing our understanding of high-energy phenomena in astrophysics, including predicting the outcome of the BNS merger and determining the central engine of GRBs.
Such independent measurements can also directly test the $M_{\rm TOV}$ and $R_{\rm TOV}$ inferred in other ways \citep[see][and references therein]{2020ApJ...904..119F, 2023arXiv230912644F}.
Additionally, since the $M_{\rm crit}-\chi$ universal relation differs between NSs and quark stars \citep{2018MNRAS.474.3557B,2019PhRvD.100d3015Z}, our method could potentially be further extended to differentiate the nature of compact stars.

Finally, we would like to discuss some potential difficulties/challenges when people apply our method to real observations.
Firstly, distinguishing between high-mass NSs and low-mass BHs based solely on GW signals is challenging.
However, there are promising methods proposed for third-generation GW detectors to mitigate this issue \citep{2020PhRvD.101j3008C, 2022ApJ...941...98B}.
Moreover, observation data from EM counterparts, such as kilonova observations, can also assist in distinguishing between the two types \citep{2020ApJ...889..171K} and in constraining mass and spin parameters \citep{2019A&A...625A.152B} when available. 
Secondly, confirming the presence of 2G BHs presents its own unique challenges.
Current mass-spin population studies for BBHs suggest that first-generation (1G) BHs typically exhibit low spin \citep{2023PhRvX..13a1048A, 2022ApJ...941L..39W, 2023arXiv230302973L}.
And binary evolution simulation also shows that the BHs in ordinary BH-NS systems have dimensionless spin parameter less than about 0.2 \citep{2023arXiv230909600X}.
If we attribute high-spin values of low-mass BHs mainly to BNS mergers, our sample remains relatively uncontaminated.
However, if we consider accretion onto low-mass 1G BHs as a contributing factor, then these additional BH populations could potentially contaminate our sample and cause biased measurements of $M_{\rm TOV}$ and $R_{\rm TOV}$ as discussed in the Results section.
In such instances, these populations should be carefully modeled and subsequently excluded.
Thirdly, the detection rate of (2G)BH-NS mergers carries a significant degree of uncertainty.
Nevertheless, some studies have indicated that certain detected GW events support the hypothesis of a (2G)BH-NS merger origin \citep{2020PhRvD.101j3036G,2021MNRAS.502.2049L,2023arXiv230709097G}, and the estimated merger rate appears promising \citep{2021PhRvD.104d3004K}. 
Fourth, the 2G BHs may not remain in the same state as they were initially formed from the collapse of SMNSs.
This is due to the fact that BNS mergers might produce GRBs, and the Blandford-Znajek mechanism \citep{1977MNRAS.179..433B} could cause the newly formed BHs to spin down.
In addition, the mass that falls into these BHs may contribute to their growth in mass.
As indicated by the results from numerical simulations \citep{2020PhRvD.101h3029F}, the shift in $\chi_{\rm BH}$ appears insignificant compared to its measurement uncertainty.
However, the mass that falls into the BH is a critical factor, and the masses of 2G BHs ($M_{\rm BH}$) need to be accurately calibrated.

\begin{acknowledgments}
We would like to thank Francesco Iacovelli for the kind help in using GWFAST.
This work is supported in part by the Project for Special Research Assistant and the Project for Young Scientists in Basic Research (No. YSBR-088) of the Chinese Academy of Sciences, 
the National Natural Science Foundation of China under grant Nos. 12233011, 11921003, 11933010, 12073080, and 12303056, 
and by the General Fund (No. 2023M733736) of the China Postdoctoral Science Foundation.
\end{acknowledgments}

\software{Scipy \citep[version 1.11.1,][\url{https://scipy.org}]{2020NatMe..17..261V}, GWFAST \citep[version 1.1.1,][\url{https://github.com/CosmoStatGW/gwfast}]{2022ApJS..263....2I}, RNS \citep[version v1.1d,][\url{https://github.com/cgca/rns}]{1995ApJ...444..306S}}

\bibliography{ms.bib}{}
\bibliographystyle{aasjournal}

\end{document}